\documentclass[preprint,review,12pt]{elsarticle}

\usepackage{amssymb}

\usepackage[nodots]{numcompress}
\usepackage{textcomp}

 \biboptions{compress}

\journal{Carbon}

\begin{document}

\newcommand{\SiO}{SiO$_2$~}
\begin{frontmatter}



\title{Molecular beam growth of graphene nanocrystals on dielectric substrates}


\author[CUPHY]{Ulrich Wurstbauer\corref{cor1}}
 \cortext[cor1]{Corresponding author. E-mail address: uliw@phys.columbia.edu (U. Wurstbauer) }

\author[CUEFRC]{Theanne Schiros}
\author[NIST]{Cherno Jaye}
\author[EXETER]{Annette S. Plaut}
\author[CUPHY,UNI]{Rui He}
\author[CUPHY]{Albert Rigosi}
\author[CUPHY]{Christopher Gutierrez}
\author[NIST]{Daniel Fischer}
\author[PU]{Loren N. Pfeiffer}
\author[CUPHY]{Abhay N. Pasupathy}
\author[CUPHY,CUAPAM]{Aron Pinczuk}
\author[CUPHY,MBE]{Jorge M. Garcia}

\address[CUPHY]{Department of Physics, Columbia University, 538 West 120th Street,  New York, NY 10027, USA}
\address[CUEFRC]{Energy Frontier Research Center, Columbia University, 530 West 120th Street, New York, NY 10027, USA}
\address[NIST]{Materials Measurement Laboratory, National Institute of Standards and Technology, 100 Bureau Drive M/S 8520, Gaithersburg, MD 20899, USA}
\address[EXETER]{School of Physics, Exeter University, Stocker Road, Exeter, EX4 4QL, UK}
\address[UNI]{Physics Department, University of Northern Iowa, Cedar Falls, Iowa 50614, USA}
\address[PU]{Electrical Engineering Department, Princeton University, Princeton, NJ 08544, USA}
\address[CUAPAM]{Department of Appied Physics and Applied Mathematics, Columbia University, 538 West 120th Street, New York, NY 10027, USA}
\address[MBE]{MBE Lab, IMM-Instituto de Microelectronica de Madrid (CNM-CSIC), Isaac Newton 8, PTM, E-28760 Tres Cantos, Madrid, Spain}

\begin{abstract}
We demonstrate the growth of graphene nanocrystals by molecular beam methods that employ a solid carbon source, and that can be used on a diverse class of large area dielectric substrates. Characterization by Raman and Near Edge X-ray Absorption Fine Structure spectroscopies reveal a sp$^2$ hybridized hexagonal carbon lattice in the nanocrystals. Lower growth rates (below 0.5 \AA/min) favor the formation of layered, larger size multi-layer graphene crystallites (up to 22~nm) on all investigated substrates. The surface morphology is determined by the roughness of the underlying substrate (RMS $\sim 8$~\AA) and graphitic monolayer steps are observed by ambient scanning tunneling microscopy.
\end{abstract}

%
%

\end{frontmatter}


\section{Introduction}
Realization of  the enormous potential of graphene in condensed matter science and nanotechnology requires use of state-of-the-art processing methods. While mechanical exfoliation of graphite offers ultra-high quality graphene flakes on arbitrary substrates, it is not suitable for creation of large-area films required for many experiments and for potential applications \cite{Novoselov-1stScience}. Thermal decomposition of SiC \cite{Berger-deHeer-JPhysChem} and chemical vapor deposition (CVD) on transition metals result in large area material \cite{Kim-NiCVD-nature-2009,Ruoff-Science2009,Reina-CVDNanoletter-2008}, but the methods are limited to specialized substrates.
\par
Molecular-beam epitaxy (MBE) enables growth of precisely tailored semiconductor structures that have greatly advanced fundamental condensed matter science \cite{LorenPfeiffer-PhysicaE-2003}. MBE growth of superb semiconductor quantum structures is facilitated  by access to suitable substrates and by availability of atomic or molecular beams from ultra-clean sources in an ultra-high vacuum (UHV) environment. The lack of a substrate with suitable interface properties for epitaxy hinders successful MBE-growth of graphene. MBE-like growth of graphene layers on arbitrary substrates would enable electrical and magnetic doping and the creation of sophisticated graphene-based heterostructures for fundamental research. Consequently, development of MBE growth of graphene on various substrates is currently an area of intense research activity \cite{AlTemimy-APL-2009,Hackley-APL-2009,Moreau-CMBEonSiC-APL-2010,Park-CMBE-APL-2010, Maeda-gasCMBE-JAPL,Tang-GrowthOnSi-PhysicaE,Jerng-JoPhysChemC-2011-GNCSapphire,Lippert-PSSB-2011-CMBE}.
\par
We report the fabrication of large area conducting graphene films by a MBE-inspired method that we call molecular-beam-growth (MBG).  In MBG the films are directly grown on the substrate surface by a carbon beam that is created from a solid carbon source in an UHV environment. We demonstrate  that MBG films of graphene nanocrystals can be grown on a diverse group of dielectric substrates.  The combination of highly controllable growth conditions and dielectric substrates produces films that do not require exfoliation for further experiments, and facilitates comprehensive in-depth characterization. 
\par
We employ Raman scattering spectroscopy, Near edge X-ray absorption fine structure spectroscopy (NEXAFS), and scanning tunneling microscopy (STM) to explore the key parameters required to obtain good quality ultra-thin graphitic films by MBG. Raman spectroscopy probes the crystallinity and overall quality of the layers, and shows that the carbon atoms are arranged in a hexagonal lattice. NEXAFS spectroscopy is used to characterize the nature and geometry of the carbon bonds and overall quality of the films, including sp$^2$:sp$^3$ ratios \cite{Stoehr-NEXAFS}. The NEXAFS data unequivocally demonstrate planar layered sp$^2$ graphitic bonds in films grown under optimized conditions. The STM measurements reveal monoatomic steps and electrical conductivity. 
\par
The quality and size of the graphene nanocrystals in the MBG films strongly depend on the growth conditions. Most notably, we find that the growth rate (GR), in particular, plays a crucial role.  If the GR is too high the sp$^2$ carbon bonds in the nanocrystals form a three-dimensional (3D) arrangement that precludes the layered structure of graphene multilayers. By lowering the GR (below 0.5\AA/min) we successfully grow layered MBG films that show clear signatures of a two-dimensional lattice of sp$^2$-carbon in NEXAFS and Raman spectroscopy. On the basis of these results we discuss the MBG growth mode of ultra-thin films of graphene nanocrystals.
\par

\section{Experimental methods}
\subsection{Molecular beam growth} 
\begin{figure}
\includegraphics[width=9cm]{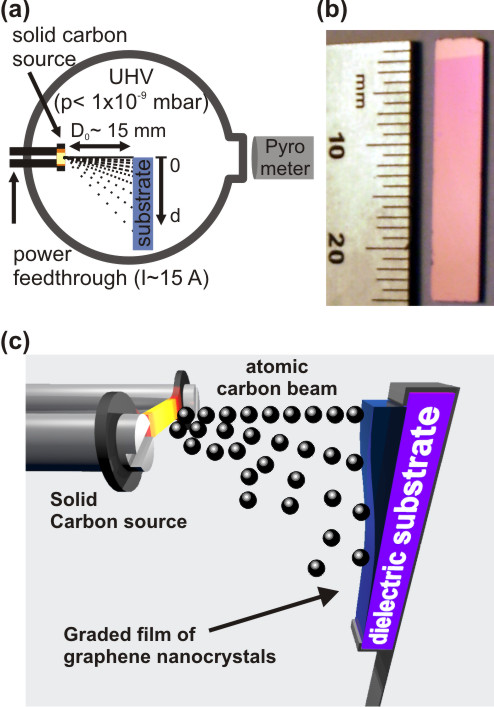}
\caption{ (a) Schematic of the growth arrangement - the solid carbon source is placed closely facing the substrate. This geometry leads to a gradient in growth rate along the long axis of the sample.  (b) Photograph of an ultra-thin graphene MBG film on a \SiO substrate. The changes in color are from the graded thickness, and the contrasting colour at the top is due to mechanical clamping of the substrate during growth. (c) Schematic of the MBE-inspired growth.}
\label{fig:setup}
\end{figure}
MBG film growth of graphene nanocrystals on dielectric substrates, such as amorphous SiO$_2$ (300~nm) and crystalline Mica is achieved in the set-up illustrated in Fig. \ref{fig:setup}. The substrates are cleaned by sonication in acetone and isopropanol prior to loading in the growth chamber. Mica samples are cleaved {\it ex-situ} and are loaded immediately into the UHV system. The UHV chamber (base pressure $\sim 6\times 10^{-10}$~mbar) incorporates a solid carbon source that is made of glassy carbon, similar to that employed for p-doping in III-V MBE \cite{Manfra-GaAsCdop-APL-2005} and other carbon-related growth \cite{JorgeG-MultilayerDeposition-SSC-2010,JorgeG-PUC-Carbon-2011}. The dimensions of the carbon source are 10~\texttimes~2.5~\texttimes~0.3 mm$^3$. The carbon source is heated by a DC current of $\sim$15~A to an operating temperature of $\sim$~2100\textcelsius~which is monitored by a Marathon MM Raytech optical pyrometer. The solid carbon source is located in close proximity to the substrate, as shown in Figs. \ref{fig:setup}(a) and (c). The substrates are heated to $\sim$~400\textcelsius~to remove adsorbed water before the growth. The maximum pressure reached during growth is $\sim 5\times 10^{-8}$~mbar. Due to the proximity of the solid carbon source, the temperature of the substrates during growth reaches $\sim$~500\textcelsius.
\par
In the growth set-up shown in Fig. 1(a), $D_0$ is the distance between the carbon source and the sample ($\sim$~15~mm) and $d$ is the position on the substrate. $\Theta_0$ is the thickness at $d=0$. In this configuration the flux of carbon atoms is relatively high at the near end of the substrate (d~=~0) and decreases significantly along the length of the substrate. The geometrical dependence of the flux is best described as a GR gradient along the length of the substrate \cite{JorgeG-MultilayerDeposition-SSC-2010}. 
The calibration of the GR is achieved by measuring the profile of a thick MBG films ($>$ 30 nm) on a \SiO substrate using an atomic force microscope or optical profilometer. The position-dependent GR$(d)$, derived from the position-dependent thickness $\Theta (d)$, is calculated according to the expression \cite{JorgeG-MultilayerDeposition-SSC-2010}
\begin{equation}
\mathrm{GR}(d)=\frac{\Theta(d)}{t}=\frac{\Theta_0/\left(1+\left(\frac{d}{D_0}\right)^2\right)^2}{t},
\end{equation}
where $t$ is the deposition time. The maximum GR (typically 1-2 \AA/min) is reached for $d=0$. As $d$ increases GR$(d)$ decreases to a minimum value of 0.1 \AA/min, or less.
\subsection{Characterization methods}
Raman spectroscopy is a widely used technique for characterization of carbon-based materials \cite{Tuninstra-Grainsize-1970,FerrariRobertson-PRB,Ferrari-Raman-PRL,Dresselhaus-Review-2010,Pimenta-Review-2007}. NEXAFS provides a direct, element-specific probe of bond type and orientation with a high surface sensitivity that enables evaluation of sp$^2$:sp$^3$-bond ratios and the degree of planarity of ultra-thin (single layer) films \cite{Stoehr-NEXAFS}. Since sp$^2$-hybridized carbon layers have unique spectral fingerprints in both Raman and NEXAFS spectroscopies, the combination of these two methods is particularly suited to probing the crystallinity, bond type and bond configurations (2D vs 3D) of the ultra-thin MBG films.
\par
For the Raman experiments a Renishaw inVia micro-Raman set-up, equipped with a movable x-y-z stage was employed. The laser power was set to less than 3~mW and was focused with a 100\texttimes~lens to a spotsize of $\sim$~0.5~$\mu$m.
\par
Carbon~1s NEXAFS measurements have been performed at the NIST beamline U7A of the National Synchrotron Light Source (NSLS). Measurements were performed in partial electron yield (PEY) mode with a grid bias of -200 V, selected to optimize the surface sensitivity of the measurement and thereby the signal from the graphene film. Angle-dependent NEXAFS was obtained by changing the angle between the incoming X-ray beam (and therefore the E-field vector) and the sample between 20\textdegree and 70\textdegree, corresponding roughly to out-of-plane and in-plane bond resonances, respectively. The reference absorption intensity (I$_0$) of the incoming X-ray beam, measured on a gold coated mesh positioned just after the refocusing optics, was measured simultaneously and used to normalize the spectra to avoid any artifacts due to beam instability. A linear background was subtracted from a region before the absorption edge (278-282~eV). Spectra were normalized by area with respect to carbon concentration using a two-point normalization: area normalization between 282 and 300~eV and a continuum normalization in the region 330-335~eV (atomic normalization).
\par
Ambient STM and atomic force microscopy (AFM), in tapping mode, have been performed to get additional insight into the surface morpholgy of the grown films.

\begin{figure*}
\includegraphics[width=\columnwidth]{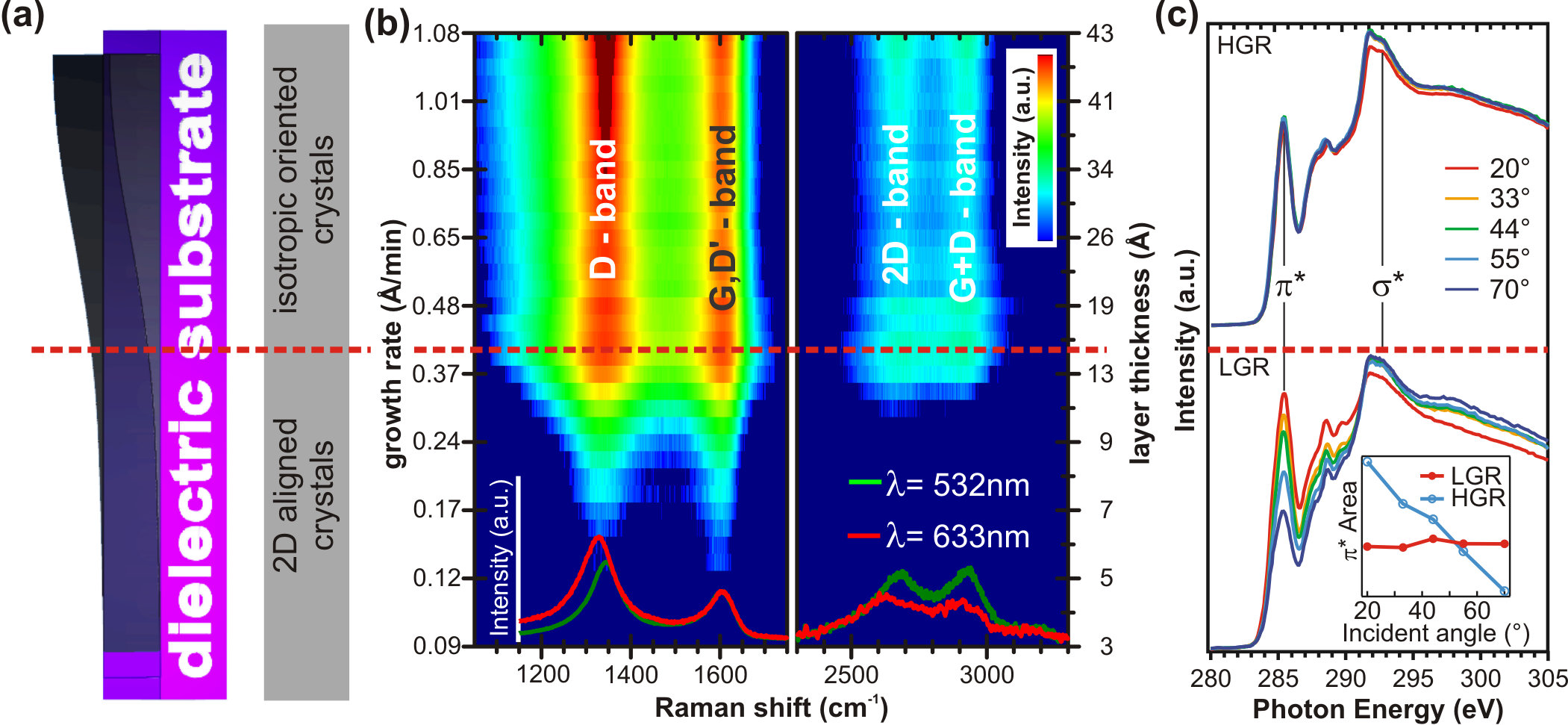}
\caption{(a) Schematic separating two distinctively different regions: High growth rates (HGR) result in isotropically arranged graphene layers. Low growth rates (LGR) are dominated by two-dimensional growth parallel to the substrate surface. (b) Color map of the Raman intensities along the GR gradient and concomitant varying growth rate. Typical features observed can be linked to graphitic material, such as the D, G, 2D(G') and G+D peaks. Inserted are two Raman traces at two excitation wavelengths at GR~=~1.08~\AA/min. (c) NEXAFS spectra showing characteristic sp$^2$ bond features: $\pi^*$ and $\sigma^*$. For HGR, no polarization dependence of the  $\pi^*$ and  $\sigma^*$ resonances is observable. For LGR, a large polarization dependence is observed indicating layered growth of the graphene crystals. The inset depicts the dependence of the integrated $\pi^*$ peak intensity on the beam orientation for both HGR and LGR.}
\label{fig:resultsGradient}
\end{figure*}
\section{Experimental results and discussion}
Figure \ref{fig:resultsGradient} shows typical Raman and NEXAFS measurements for a MBG film grown on a 300~nm-thick \SiO layer on Si. As discussed below, these results show, unambiguously, the formation of graphene nanocrystals along the whole gradient and reveal that the grown layers  have different structural properties depending on the GR. Two main regions with very distinct characteristics can be identified. The dashed line in Fig. \ref{fig:resultsGradient} marks the border between these two regions, corresponding to high GR (upper half) and low GR (lower half), respectively. 
\subsection{MBG films on amorphous \SiO substrates}
\subsubsection{Raman spectroscopy}
Characteristic Raman signatures of optical phonons for graphite \cite{Tuninstra-Grainsize-1970, FerrariRobertson-PRB, Ferrari-Raman-PRL, Dresselhaus-Review-2010, Pimenta-Review-2007, Ferrari-SolidStateComm2007, Nemanich-GraphitePRB-1979, MartinsFerreira-RamanSpectraPRB-2010, Yan-2Dgating-PRL-2007} are observed along the GR gradient, as displayed in the color plot of Fig. \ref{fig:resultsGradient} (b). The band at $\sim$~1600~cm$^{-1}$ results from superposition of the G and the D' modes. The G mode is a long wavelength optical phonon originating from in-plane bond-stretching motion of pairs of sp$^2$ hybridized carbon atoms. The D' mode is induced by disorder and requires intra-valley electron-phonon scattering. The D mode at $\sim$~1344~cm$^{-1}$, which requires the presence of six-fold aromatic rings, is induced by disorder, such as edges or atomic defects. The bands resolved at higher Raman shifts are also well known: The 2D mode at $\sim$~2700~cm$^{-1}$ (a.k.a. G'), the G+D band slightly below $\sim$~3000~cm$^{-1}$ and a third one at $\sim$~3200~cm$^{-1}$ matching the energy of the G+D' mode. We observe the latter, but it is not discernible in the color plot of Fig. \ref{fig:resultsGradient} (b).
\par
The intensity of all Raman features decreases with decreasing GR (film thickness), while the relative intensities of D and G bands vary with the GR: For higher GR (upper part of Fig. \ref{fig:resultsGradient}), the D mode is more intense than the G mode (see left panel of Fig. \ref{fig:resultsGradient} (b)). In addition, at higher GR a significant Raman intensity is observed between the G and D modes, which originates from the presence of disordered carbon bonds \cite{FerrariRobertson-PRB, Ferrari-SolidStateComm2007}. At lower GR (below the dashed line of Fig. \ref{fig:resultsGradient}) the peak intensity ratio $I(D)/I(G)$ greatly diminishes and the G and D bands become better resolved due to the Raman intensity between those two modes decreasing drastically. Both observations point to a larger crystal size and a higher crystal quality for lower growth rates.
\par
The two Raman spectra (GR~=~1.08~\AA/min), shown at the bottom of Fig. \ref{fig:resultsGradient} (b), reveal that by changing the excitation laser wavelength from $\lambda=532$~nm (green trace) to $\lambda=633$~nm (red trace) there is a clear redshift in the positions of the D and 2D Raman bands. In crystalline graphene layers, such a redshift arises from the wave-vector dispersion of the optical phonons. The size of the frequency shift that we observe is comparable to those reported for graphite and graphene \cite{FerrariRobertson-PRB, Ferrari-Raman-PRL,Dresselhaus-Review-2010,Pimenta-Review-2007,Ferrari-SolidStateComm2007,ThomsenReich-PRL-2D-2000}. The observed energy dispersion of the MBG films provides further evidence of crystallinity.

\subsubsection{NEXAFS spectroscopy}
The two growth regions have the distinct NEXAFS signatures, as shown in Fig. \ref{fig:resultsGradient} (c). In both regions we find spectral fingerprints of sp$^2$-hybridized carbon \cite{Stoehr-NEXAFS}: strong peaks at 285.4~eV and 292.0~eV that correspond to excitation of a carbon 1s core electron to the unoccupied  $\pi^*$ and  $\sigma^*$ orbitals, respectively. The sharpness of the NEXAFS features indicates a well-defined bonding environment and long-range periodic order in the electronic structure. The  $\sigma^*$ fine structure, in particular, is specifically characteristic of graphite, and includes a sharp onset due to an excitonic core hole-valence state interaction and the broader  $\sigma^*$ peak at $\sim$~1 eV higher photon energy due to more delocalized $\sigma^*$ states \cite{Bruehenwiler-PRL-1995}. Thus, the NEXAFS spectra unequivocally prove the formation of sp$^2$ bonds between carbon atoms in the MBG films. 
\par
NEXAFS is also sensitive to substrate-relative bond-orientations \cite{Stoehr-NEXAFS}. Being governed by the transition dipole matrix element between a core electron and an unoccupied orbital above the Fermi level, the NEXAFS intensity depends upon the angle between the electric field vector of the incoming X-ray beam and the molecular orbitals in the system (see inset of Fig. \ref{fig:PolDepNEXAFS}). Hence, we directly probe the degree of bond anisotropy in the sp$^2$ films by changing the angle of the incident x-ray beam from near parallel (20\textdegree) to near perpendicular (70\textdegree) to the substrate; the E-field vector is perpendicular to the beam axis. 
\par
For higher GR (upper half of Fig. \ref{fig:resultsGradient} (c)) no angular dependence of the NEXAFS resonances is observed, indicating a fairly isotropic arrangement of sp$^2$ bonds. 
In striking contrast, the NEXAFS intensity becomes strongly dependent on incident angle at lower GR (lower half of Fig. \ref{fig:resultsGradient} (c)). The intensity of the $\pi^*$ ($\sigma^*$) peak is at its maximum (minimum) at 20\textdegree~and minimum (maximum) at 70\textdegree~ incidence, indicating highly oriented planar C=C bonds parallel to the substrate surface. Here, the sp$^2$ carbon layers grow in a two-dimensional plane. The ability to grow sp$^2$ carbon layers well aligned to the plane of the substrate, and the presence of two regions with distinctly different degrees of bond anisotropy is emphasized by the inset of Fig. \ref{fig:resultsGradient} (c), which plots the area of the $\pi^*$ peak as a function of the incident angle for the two regions.
\par
The homogeneity of the material throughout the volume can be probed with NEXAFS by varying a bias voltage applied to the sample. By changing the voltage from -250~V to -50~V, the depth within the carbon film from which detected electrons are emitted can be tuned from about 1~nm to about 7~nm (maximum film thickness $\Theta_0 < 3.5$~nm). The higher voltage allows detection of electrons only from the near surface-region. Since we do not observe a bias-dependency of the spectral features, we conclude that the films are homogenous throughout the volume. This excludes the possibility of initial formation of a planar film in the isotropic region of the films followed by accumulation of defects as the film thickness is increased.

\subsection{Angle-dependent NEXAFS: results on various substrates}
\begin{figure}
\includegraphics[width=9cm]{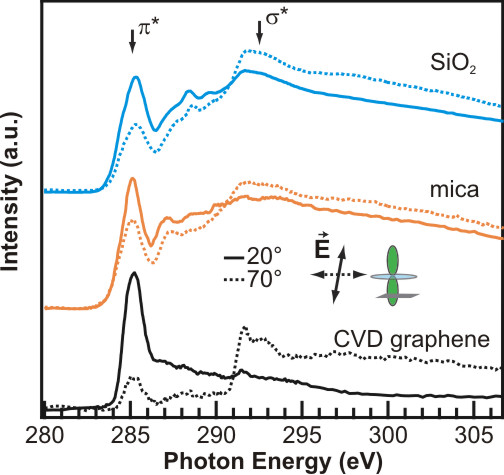}
\caption{Angle dependent NEXAFS measurements in the low-growth-rate region. The similarity of MBG graphene grown on \SiO and mica with CVD graphene, including the energy position and angular dependence of the NEXAFS features, is clearly observable. Incident angles of 20\textdegree~ and 70\textdegree~ correspond roughly to out-of-plane and in-plane polarizations, as shown schematically in the inset.}
\label{fig:PolDepNEXAFS}
\end{figure}
\par
Figure \ref{fig:PolDepNEXAFS} compares angle-dependent NEXAFS spectra for few-layer ($\sim$2~nm) MBG graphene on \SiO and on mica, with  a single high-quality graphene layer grown on Cu foil by CVD. The NEXAFS spectra from the MBG and CVD grown graphene are very similar. In fact, aside from a slightly weaker angular dependence of the MBG films, the main difference between the MBG and CVD spectra is the intensity in-between the $\pi^*$ and the $\sigma^*$ resonances, which is due to C-O and C-H bonds (a resonance due to an interlayer state in few-layer graphene also appears in this region) \cite{Rusz-XrayHBNGraphene-PRB-2010,NEXAFS-CO-CH-bonds2009,NEXAFS-Coffman}.
\par
This intensity between the $\pi^*$ and the $\sigma^*$ resonances can be explained by the larger number of dangling bonds available at the grain boundary of the MBG nanocrystals, due to their smaller grain size compared to those in the CVD samples. These are readily saturated by oxygen and hydrogen bonds. These bonds tend to distort the planarity of graphene films and therefore may also explain the suppressed angular dependence of the NEXAFS data for the MBG films compared to CVD graphene.
\par
Importantly, no features associated with sp$^3$ carbon-carbon bonds are observed in the NEXAFS data.

\subsection{NEXAFS results from thick MBG films}
\begin{figure}
\includegraphics[width=9cm]{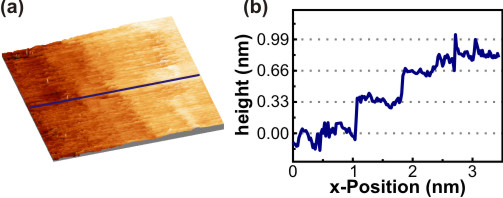}
\caption{Orientation independent NEXAFS of bulk material measured near the magic angle (50\textdegree). The spectra from HOPG and a 50~nm-thick layer grown by MBG show the same detailed structure. In contrast to that, the glassy carbon trace shows a strongly suppressed sp$^2$-$\pi^*$ resonance, a shift and broadening of the sp$^2$-$\sigma^*$ and a well pronounced peak at $\sim$~289~eV, which is associated with the $\sigma^*$ resonance of sp$^3$-carbon.}
\label{fig:NEXFASthickStuff}
\end{figure}
To further investigate the impact of disorder, we compare NEXAFS data from a thick MBG film ($\Theta_0=54.4$~nm), both with that from glassy carbon (used as carbon source) and from highly-ordered pyrolytic graphite (HOPG). This is shown in Fig. \ref{fig:NEXFASthickStuff}.
\par
While the NEXAFS spectrum of the MBG film is very similar to that of HOPG, distinct differences are observed from the glassy-carbon spectrum, which has significant sp$^3$ content. In particular, the sp$^2-\pi^*$ and $\sigma^*$ peaks are strongly suppressed and the sp$^2-\sigma^*$ peak is significantly broadened. An onset and a peak centered around 289 eV appears due to the sp$^3-\sigma^*$ absorption edge of diamond and a C-H resonance \cite{NEXAFS-CO-CH-bonds2009,NEXAFS-Coffman}.
\par
In contrast, the thick MBG film and the HOPG traces possess the spectral signatures of sp$^2$ bonds, as already discussed. HOPG has better long range periodic ordering,  as is evidenced by the sharpness of the $\sigma^*$ resonance. As in Fig. \ref{fig:PolDepNEXAFS}, the MBG films show some C-H and C-O bonds at the grain boundaries of the nanocrystals, as well as non-uniform bonding between the differently oriented graphene nanocrystals in three dimensions, giving rise to the intensity between the sp$^2-\sigma^*$ and the $\pi^*$ resonances (indicated by the arrow in Fig. \ref{fig:NEXFASthickStuff}).\\
Thus, the thick MBG films are indeed structurally very similar to HOPG and consist of graphite nanocrystals.

\subsection{Scanning tunneling microscopy}
\begin{figure}
\includegraphics[width=9cm]{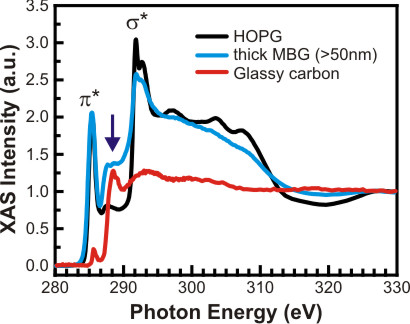}
\caption{Ambient STM measurements reveal terraces with a step height of roughly 0.33~nm as shown in (b), which shows the line profile along the blue line in (a).}
\label{fig:STM}
\end{figure}
As expected for sp$^2$ bonded carbon, the MBG films show electrical conductivity at room temperature. Preliminary 4-probe transport measurements reveal a sheet resistivity of a few k$\Omega$; sufficient conductivity for STM measurements. Figure \ref{fig:STM} (a) shows a 3-dimensional ambient STM topography of a MBG film on a mica substrate (the size of the image is 4\texttimes 4~nm$^2$). The measurement was done at the crossover to the low GR region ($\sim 0.5$~\AA/min).   Several flat terraces are observed. A line profile, along the blue line in \ref{fig:STM} (a), reveals 0.33~nm high steps, as shown in Fig. \ref{fig:STM} (b). These step-heights are comparable to the interlayer distance in graphite \cite{hembacher-Stepheihgt-PNAS-2003}, as one would expect in graphene multilayers. The surface roughness is dominated by the roughness of the underlying substrate and has a RMS value of $\sim 8$~\AA. This is in agreement with tapping-mode AFM measurements of larger areas showing RMS roughness values of 0.2 to 0.3~nm across the whole GR gradient.

\section{Analysis of Raman spectra: nanocrystal size}
\label{ref:cyrstalsize}
\begin{figure}
\includegraphics[width=9cm]{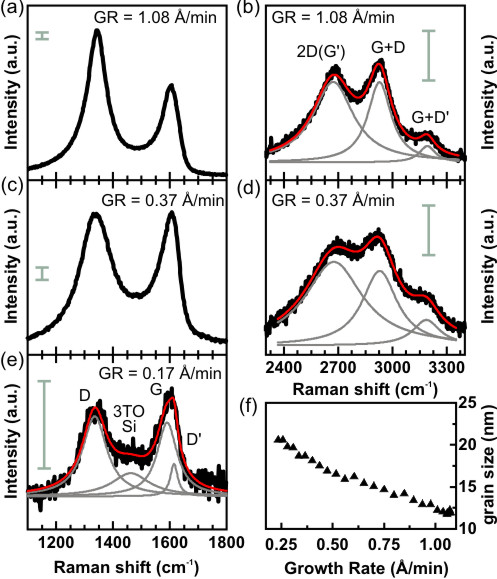}
\caption{Micro Raman spectra of MBG graphene nanocrystals on amorphous \SiO measured at various GRs: (a and c) the D and G modes. (b and d) the G'(2D), G+D and G+D' modes. The spectra in (c) and (d) are typical for the crossover region. (e) a micro-Raman spectrum acquired at low GR: The D'-mode intensity decreases and the region between the D and G bands changes to show an additional peak attributed to a 3TO-Si phonon. (f) The crystal grain size estimated from the ratio of the D and the G modes. Reduction of the GR leads to an increase of grain size. Scale bars represent one arbitrary unit.}
\label{fig:RamanDetails}
\end{figure}
Detailed analysis of Raman lineshapes enables estimates of the crystallite sizes. Typical Raman spectra of MBG nanocrystals grown on \SiO are shown in Fig. \ref{fig:RamanDetails}. A (linear) background subtraction between 1900~cm$^{-1}$ and 2300~cm$^{-1}$, has been applied.
\par
Figure \ref{fig:RamanDetails} (a)-(d) show results at high GRs. The blue shift of the G band to $\sim$~1600~cm$^{-1}$ (from $\sim$~1585~cm$^{-1}$ in graphite) seen in  Figs. \ref{fig:RamanDetails} (a) and (c) is attributed to the unresolved superposition of the G and D' Raman modes \cite{FerrariRobertson-PRB,Ferrari-SolidStateComm2007,Pimenta-Review-2007}. The Raman intensity between the D and G band is tentatively interpreted as from disorder at the grain boundaries. In addition to the 2D (G') band at ($\sim~2672\mathrm{~cm}^{-1}$) we also observe two second-order bands at higher Raman shifts (Figs. \ref{fig:RamanDetails} (b) and (d)): G+D at $\sim$~2928~cm$^{-1}$ and G+D' at $\sim$~3202~cm$^{-1}$. These allow a calculation of the energy shifts of the G and D' modes, which thereby are found to be at 1584~cm$^{-1}$ and 1618~cm$^{-1}$, respectively. All these values are in very good agreement with previous reports for such graphene-like systems \cite{FerrariRobertson-PRB,Ferrari-Raman-PRL,Dresselhaus-Review-2010,Pimenta-Review-2007,Ferrari-SolidStateComm2007,Nemanich-GraphitePRB-1979,MartinsFerreira-RamanSpectraPRB-2010,Yan-2Dgating-PRL-2007}.
\par
Figure \ref{fig:RamanDetails} (e) shows results at lower GRs: Here the G mode redshifts to $\sim$~1585~cm$^{-1}$, indicating that the contribution of the D' band is reduced, most likely as a consequence of a larger nanocrystal size. In addition, the D-mode intensity is reduced relative to the G mode and the intensity between those two modes decreases. Four Lorentzians, corresponding to the three graphene optical phonon frequencies: D, G and D'; and a fourth one related to the 3TO Si phonon (at 1450~cm$^{-1}$), reproduce the data \cite{Zwick_PRB-Si3TO}.
\par
The intensity ratio I(D)/I(G), provides an estimate of the crystallite dimensions \cite{Tuninstra-Grainsize-1970,FerrariRobertson-PRB,Pimenta-Review-2007,Ferrari-SolidStateComm2007,PimentaCancado-APLGrainSize2006}. The graph of Figure \ref{fig:RamanDetails} (f) reveals an unambiguous trend: that the grain size increases up to 22~nm on reducing the GR. This result is consistent with the reduced Raman intensity between the D and G lines in Fig. \ref{fig:RamanDetails} (e).

\section{Discussion}
These experimental findings, suggest a preliminary picture of how these MBG films are formed, which indicates that fundamentally different atomic arrangements are caused by variations in the growth rate. It is worth emphasizing that this phenomenological description is based not only on the samples presented here, but also on sets of samples grown at different maximum GRs (2.3~\AA/min to  $< 0.75$~\AA/min), various substrate temperatures (380\textcelsius~to 940\textcelsius) and different maximum film thicknesses ($\Theta_0 \sim$ 3.3~nm to 15.6~nm as defined in Eqn. 1). 
\par
After controlling for all the other variables in the growth, our main finding is that lower fluxes of carbon atoms create graphene layers that are more aligned with the substrate surface. This is clearly seen for example in Figure 2 where the differing orientations are observable on the same sample.  The upper region, being closer to the solid carbon source, is subject to a higher rate of incident carbon atoms. The NEXAFS and Raman spectra from this upper region reveal isotropic orientations of the sp$^2$-bonds. In the lower region, due to its increased distance from the carbon source, and thereby its lower growth rate, we find graphene multilayers which are aligned parallel to the substrate surface.
\par
The diversity of carbon chemistry implies that bonding between individual carbon atoms on a surface can proceed via several reaction pathways depending on the details of the environment. With so many reaction paths, there are many opportunities for carbon bonding, which seems to occur very fast. Thus, film growth can be understood as a dynamic process which strongly depends on the number of available carbon atoms that find themselves in close proximity within the time it takes to form a bond. In the limit of a large flux of atoms, each carbon atom has not enough time to find an energetically-optimized location before forming a covalent bond. This leads to the formation of bonds along random directions, resulting in small randomly-aligned graphene nanocrystals. This growth can be regarded as quasi-3D and has similarities to Volmer-Weber (island) growth. Conversely, layered material can form, if the impinging carbon atoms have more time to find energetically-favored locations and bond on a larger graphene edge. For a given substrate temperature, this reduction in the number of available carbon atoms is realized by reducing the GR. This layered growth is strongly analogous to two-dimensional Frank-van der Merwe  (layer-by-layer) growth, but further investigations are needed to confirm this. A roughness analysis reveals a smooth growth front, independent of the film thickness, for low GRs, which is typical of the Frank-van der Merwe growth mode.
\par
Increasing the substrate temperature might be expected to have a similar effect as the reduction of the GR due to a higher diffusion rate of the atoms on the surface. However, our findings show that growths at higher substrate temperatures result in sp$^3$ bonds being present in the films. We speculate that a thermally-induced, chemical interaction of carbon atoms with the substrate is activating the sp$^3$ formation. In addition, the diffusion of carbon atoms on the surface may be limited by roughness of the \SiO substrate. 
\par
In order to exclude the influence of film thickness, we have also studied samples with films of different thicknesses that were grown by varying the carbon flux. The results are in agreement with the bias-dependent NEXAFS measurements discussed above (Section 3.1.2). Independent of the film thickness, high GRs ($\sim 1-2$~\AA/min) result in isotropic material (3D growth) throughout the film volume, whereas low GRs ($< 0.5$~\AA/min) produce sp$^2$ layered films parallel to the sample surface with a concomitant increase in the size of the graphene nanocrystals.

\section{Conclusion}
We have investigated non-epitaxial growth of graphene on insulating substrates by using a molecular beam of carbon atoms and have explored, in particular, the growth rate required to obtain good quality, ultra-thin films by MBG.
\par
Using NEXAFS and Raman spectroscopy we have shown that growth rate is a crucial parameter for two-dimensional (layered) growth of graphene crystals, as it strongly influences the alignment of the sp$^2$-bonds. NEXAFS spectra for high growth rates reveal isotropic orientation of the sp$^2$-bonds. This growth can be regarded as quasi-3D.
\par
On reducing the growth rate we concomitantly increase crystallite size to $\sim$~22~nm and align the graphene multilayer-crystals parallel to the substrate. The reduction of grain boundaries manifests as reduced Raman scattering intensity between the D and G bands and anisotropy in the bond-orientations in angle-dependent NEXAFS measurements.
\par
Typical MBG parameters (growth rate, substrate temperature, surface mobility), and the MBG setup itself, offer a wide parameter space in which to explore the growth of a range of layered materials with van-der Waals coupling between the layers. At the same time MBG allows for the potential growth of heterostructures based on these layered materials.  We expect that even the use of smoother and more inert substrates, like hexagonal boron nitride, could greatly improve the crystal quality.
\section*{Acknowledgements}
The authors thank Philip Kim, Ursula Wurstbauer and Young-Jun Yu for fruitful discussions and help.
This work is supported by ONR (N000140610138 and Graphene Muri), AFOSR (FA9550-11-1-0010), EFRC Center for Re-Defining Photovoltaic Efficiency through Molecule Scale Control (award DE-SC0001085), NSF (CHE-0641523), NYSTAR, CSIC-PIF (200950I154), Spanish CAM (Q\&C Light (S2009ESP-1503), Numancia 2 (S2009/ENE-1477)) and Spanish MICINN (TEC2008-06756-C03-01,TEC2011-29120-C05-04, MAT2011-26534). Certain commercial names are presented in this manuscript for the purpose of illustration and do not constitute an endorsement by the National Institute of Standards and Technology.












\end{document}